\documentclass[preprint,12pt]{article}

\usepackage{amsmath,amssymb,array,calc,rotating,epsfig,psfrag, amscd, datetime}

\usepackage{color}
\usepackage[
      colorlinks=true,
      urlcolor=blue,    
      filecolor=blue,     
      citecolor=red,
      pdfstartview=FitV,
       bookmarksopen=true    
      ]{hyperref}
\usepackage[left=2.5cm,top=2.5cm,right=2.5cm,nohead]{geometry}

\newcommand{\nc}{\newcommand}

\definecolor{cardinal}{rgb}{0.6,0,0}
\definecolor{darkgreen}{rgb}{0,0.5,0}
\definecolor{golden}{rgb}{0.92, 0.7, 0}
\definecolor{midnight}{rgb}{0, 0, 0.5}
\definecolor{darkblue}{rgb}{0.2, 0, 0.8}

\nc{\ra}{\rightarrow} 
\nc{\lra}{\leftrightarrow} 
\nc{\Ra}{\Rightarrow} 
\nc{\LRa}{\Leftightarrow} 
\nc{\blp}{{\big (}}
\nc{\brp}{{\big )}}
\nc{\Blp}{{\Big (}}
\nc{\Brp}{{\Big )}}
\nc{\bglp}{{\bigg (}}
\nc{\bgrp}{{\bigg )}}
\nc{\Bglp}{{\Bigg (}}
\nc{\Bgrp}{{\Bigg )}}
\nc{\slb}{{\rm [}}
\nc{\srb}{{\rm ]}}
\nc{\bslb}{{\rm \big [}}
\nc{\bsrb}{{\rm \big ]}}
\nc{\Bslb}{{\rm \Big [}}
\nc{\Bsrb}{{\rm \Big ]}}

\def\al{\alpha}

\def\eps{\epsilon}
\nc{\veps}{\varepsilon}
\def\gam{\gamma}

\def\lam{\lambda}
\def\om{\omega}

\nc{\vphi}{\varphi}
\def\tha{\theta}

\def\sig{\sigma}

\def\Gam{\Gamma}

\def\Lam{\Lambda}
\def\Om{\Omega}
\def\Sig{\Sigma}

\def\coeff#1#2{\relax{\textstyle {#1 \over #2}}\displaystyle}

\nc{\myvspace}{\rule[-1em]{0pt}{2.5em}}
\nc{\bea}{\begin{eqnarray}}
\nc{\eea}{\end{eqnarray}}
\nc{\be}{\begin{equation}}
\nc{\ee}{\end{equation}}
\nc{\barr}{\begin{array}}
\nc{\earr}{\end{array}}
\nc{\co}{{\cal o}}

\nc{\cA}{{\cal A}}
\nc{\cB}{{ \cal B}}

\def\cD{{\cal D}}

\nc{\cF}{{\cal F}}
\nc{\cG}{{\cal G}}

\def\cI{{\cal I}}

\def\cK{{\cal K}}
\nc{\cL}{{\cal L}}
\nc{\cM}{{\cal M}}

\def\cN{{\cal N}}

\def\cO{{\cal O}}

\nc{\cQ}{{\cal Q}}
\nc{\cR}{{\cal R}}
\def\cS{{\cal S}}

\def\cV{{\cal V}}
\def\cV{{\cal V}}
\def\cW{{\cal W}}

\def\cZ{{\cal Z}}
\nc{\cQd}{\cQ^{\dagger}}
\nc{\cRd}{\cR^{\dagger}}
\nc{\BB}{{\mathbb B}}
\nc{\CC}{{\mathbb C}}
\nc{\DD}{{\mathbb D}}
\nc{\EE}{{\mathbb E}}
\nc{\FF}{{\mathbb F}}
\nc{\GG}{{\mathbb G}}
\nc{\HH}{{\mathbb H}}
\nc{\JJ}{{\mathbb J}}
\nc{\MM}{{\mathbb M}}
\nc{\RR}{{\mathbb R}}
\nc{\PP}{{\mathbb P}}
\nc{\QQ}{{\mathbb Q}}
\nc{\UU}{{\mathbb U}}
\nc{\ZZ}{{\mathbb Z}}
\nc{\calone}{{\mathbb 1}}
\nc{\half}{\coeff{1}{2}}
\nc{\quarter}{\coeff{1}{4}}
\nc{\del}{\partial}

\nc{\delbar}{\bar\partial}
\nc{\thalf}{\frac{t}{2}}
\nc{\Spin}{\operatorname{Spin}}
\nc{\SO}{\operatorname{SO}}

\nc{\Sp}{{\rm Sp}}
\nc{\com}[2]{{ \left[ #1, #2 \right] }}
\nc{\acom}[2]{{ \left\{ #1, #2 \right\} }}
\nc{\rr}{\rightarrow}
\nc{\p}{\partial}
\nc{\LT}{{\LL_\T}}
\nc{\Tr}{{\rm Tr}}
\nc{\tr}{{\rm tr}}
\nc{\Adag}{A^{\dagger}}
\nc{\AdagI}{A^{\dagger I}}
\nc{\AdagJ}{A^{\dagger J}}
\nc{\AdagK}{A^{\dagger K}}
\nc{\AdagL}{A^{\dagger L}}
\nc{\AdagM}{A^{\dagger M}}
\nc{\Bdag}{B^{\dagger}}
\nc{\BdagI}{B^{\dagger}_I}
\nc{\BdagJ}{B^{\dagger}_J}
\nc{\BdagK}{B^{\dagger}_K}
\nc{\BdagL}{B^{\dagger}_L}
\nc{\BdagM}{B^{\dagger}_M}
\nc{\Cdag}{C^{\dagger}}
\nc{\CdagI}{C^{\dagger I}}
\nc{\CdagJ}{C^{\dagger J}}
\nc{\CdagK}{C^{\dagger K}}
\nc{\Ddag}{D^{\dagger}}
\nc{\DdagI}{D^{\dagger I}}
\nc{\DdagJ}{D^{\dagger J}}
\nc{\DdagK}{D^{\dagger K}}
\nc{\bva}{\breve{a}}
\nc{\bvb}{\breve{b}}
\nc{\bvc}{\breve{c}}
\nc{\bvd}{\breve{d}}
\nc{\bve}{\breve{e}}
\nc{\bvf}{\breve{f}}
\nc{\bvg}{\breve{g}}
\nc{\bvh}{\breve{h}}
\nc{\bvi}{\breve{i}}
\nc{\bvj}{\breve{j}}
\nc{\bvk}{\breve{k}}
\nc{\bvl}{\breve{l}}
\nc{\bvm}{\breve{m}}
\nc{\bvn}{\breve{n}}
\nc{\bvo}{\breve{o}}
\nc{\bvp}{\breve{p}}
\nc{\brvq}{\breve{q}}
\nc{\bvr}{\breve{r}}
\nc{\bvs}{\breve{s}}
\nc{\bvt}{\breve{t}}
\nc{\bvu}{\breve{u}}
\nc{\bvv}{\breve{v}}
\nc{\bvw}{\breve{w}}
\nc{\bvx}{\breve{x}}
\nc{\bvy}{\breve{y}}
\nc{\bvz}{\breve{z}}

\nc{\bvA}{\breve{A}}
\nc{\bvB}{\breve{B}}
\nc{\bvC}{\breve{C}}
\nc{\bvD}{\breve{D}}
\nc{\bvE}{\breve{E}}
\nc{\bvF}{\breve{F}}
\nc{\bvG}{\breve{G}}
\nc{\bvH}{\breve{H}}
\nc{\bvI}{\breve{I}}
\nc{\bvJ}{\breve{J}}
\nc{\bvK}{\breve{K}}
\nc{\bvL}{\breve{L}}
\nc{\bvM}{\breve{M}}
\nc{\bvN}{\breve{N}}
\nc{\bvO}{\breve{O}}
\nc{\bvP}{\breve{P}}
\nc{\bvQ}{\breve{Q}}
\nc{\bvR}{\breve{R}}
\nc{\bvS}{\breve{S}}
\nc{\bvT}{\breve{T}}
\nc{\bvU}{\breve{U}}
\nc{\bvV}{\breve{V}}
\nc{\bvcV}{\breve{\cV}}
\nc{\bvW}{\breve{W}}
\nc{\bvX}{\breve{X}}
\nc{\bvY}{\breve{Y}}
\nc{\bvZ}{\breve{Z}}

\nc{\ul}[1]{{\underline{#1}}}
\nc{\tal}{\widetilde{\alpha}}
\nc{\tbeta}{\widetilde{\beta}}
\nc{\ttha}{\tilde{\theta}}
\nc{\ttau}{\tilde{\tau}}
\nc{\tTha}{\tilde{\Theta}}
\nc{\tphi}{\tilde{\phi}}
\nc{\tsig}{\tilde{\sig}}
\nc{\tom}{\widetilde{\om}}
\nc{\tOm}{\widetilde{\Om}}
\nc{\tlam}{\widetilde{\lam}}
\nc{\tLam}{\tilde{\Lam}}
\nc{\tSig}{\widetilde{\Sig}}
\nc{\tPhi}{\tilde{\Phi}}
\nc{\tPhibar}{\ol{\tPhi}}
\nc{\tPi}{\widetilde{\Pi}}
\nc{\tpsi}{\widetilde{\psi}}
\nc{\tPsi}{\tilde{\Psi}}
\nc{\tgam}{\widetilde{\gam}}
\nc{\tGam}{\widetilde{\Gam}}
\nc{\tzeta}{\tilde{\zeta}}
\nc{\tZeta}{\tilde{\Zeta}}
\nc{\teta}{\widetilde{\eta}}
\nc{\teps}{\tilde{\eps}}
\nc{\tveps}{\tilde{\veps}}
\nc{\tEta}{\tilde{\Eta}}
\nc{\tchi}{\tilde{\chi}}
\nc{\tChi}{\tilde{\Chi}}
\nc{\txi}{\tilde{\xi}}
\nc{\tXi}{\widetilde{\Xi}}
\nc{\tnu}{\tilde{\nu}}
\nc{\tmu}{\tilde{\mu}}

\nc{\ta}{\tilde a}
\nc{\tb}{\tilde b}
\nc{\tc}{\tilde c}
\nc{\te}{\tilde e}
\nc{\tf}{\widetilde f}
\nc{\tg}{\widetilde g}
\nc{\ti}{\tilde i}
\nc{\tj}{\tilde j}
\nc{\tk}{\widetilde k}
\nc{\tl}{\tilde l}
\nc{\tm}{\widetilde m}
\nc{\tn}{\tilde n}
\nc{\tp}{\tilde{p}}
\nc{\tq}{\widetilde{q}}
\nc{\trr}{{\tilde r}}
\nc{\ts}{{\tilde s}}
\nc{\tu}{{\tilde u}}
\nc{\tv}{{\tilde v}}
\nc{\tw}{{\tilde w}}
\nc{\tx}{{\tilde x}}
\nc{\ty}{{\tilde y}}
\nc{\tz}{\tilde z}
\nc{\tA}{{\widetilde A}}
\nc{\tAbar}{{\ol \tA}}
\nc{\tB}{{\widetilde B}}
\nc{\tC}{{\widetilde C}}
\nc{\tD}{{\widetilde D}}
\nc{\tE}{{\widetilde E}}
\nc{\tF}{{\widetilde F}}
\nc{\tG}{{\widetilde G}}
\nc{\tcG}{{\widetilde \cG}}
\nc{\tH}{{\widetilde H}}
\nc{\tI}{{\widetilde I}}
\nc{\tcI}{{\widetilde \cI}}
\nc{\tJ}{{\widetilde J}}
\nc{\tJbar}{{\ol {\tilde J}}}
\nc{\tK}{{\widetilde K}}
\nc{\tL}{{\widetilde L}}
\nc{\tcL}{{\widetilde \cL}}
\nc{\tcLbar}{{\ol \tcL}}
\nc{\tM}{{\widetilde M}}
\nc{\tN}{{\widetilde N}}
\nc{\tcN}{{\widetilde \cN}}
\nc{\tP}{{\widetilde P}}
\nc{\tQ}{{\widetilde Q}}
\nc{\tR}{{\widetilde R}}
\nc{\tS}{\widetilde{S}}
\nc{\tT}{\widetilde{T}}
\nc{\tU}{\widetilde{U}}
\nc{\tUU}{\widetilde{\UU}}
\nc{\tV}{\widetilde{V}}
\nc{\tcV}{\widetilde{\cV}}
\nc{\tcVbar}{\ol{\widetilde{\cV}}}
\nc{\tW}{\widetilde{W}}
\nc{\tcF}{\widetilde{{\cal F}}}
\nc{\tX}{\widetilde{X}}
\nc{\tY}{\widetilde{Y}}
\nc{\tcZ}{\tilde{\cZ}}
\nc{\tcZbar}{\ol{\tcZ}}

\nc{\ha}{\hat a}
\nc{\hb}{\hat b}
\nc{\hc}{\widehat c}
\nc{\hd}{\widehat d}
\nc{\he}{\widehat e}
\nc{\hf}{\widehat f}
\nc{\hg}{\widehat g}
\nc{\hh}{\widehat h}
\nc{\hm}{\widehat m}
\nc{\hn}{\widehat n}
\nc{\hp}{\widehat p}
\nc{\hq}{\widehat q}
\nc{\hr}{\widehat r}
\nc{\hs}{\widehat s}
\nc{\hv}{\widehat v}
\nc{\hw}{\widehat w}
\nc{\hx}{\widehat x}
\nc{\hy}{\widehat y}
\nc{\hz}{\widehat z}
\nc{\zhat}{\hat z}
\nc{\hA}{\widehat{A}}
\nc{\hB}{\widehat{B}}
\nc{\hC}{\widehat{C}}
\nc{\hD}{\widehat{D}}
\nc{\hE}{\widehat{E}}
\nc{\hF}{\widehat{F}}
\nc{\hcF}{\widehat{\cF}}
\nc{\hG}{\widehat{G}}
\nc{\hcG}{\widehat{\cG}}
\nc{\hH}{\widehat{H}}
\nc{\hI}{\widehat{I}}
\nc{\hcI}{\widehat{\cI}}
\nc{\hJ}{\widehat{J}}
\nc{\hK}{\widehat{K}}
\nc{\hL}{\widehat{L}}
\nc{\hcL}{\widehat{\cL}}
\nc{\hM}{\widehat M}
\nc{\hcM}{\widehat{\cM}}
\nc{\hN}{\widehat{N}}
\nc{\hO}{\widehat{O}}
\nc{\hcO}{\widehat{\cO}}
\nc{\hP}{\widehat{P}}
\nc{\hQ}{\widehat{Q}}
\nc{\hcQ}{\widehat{\cQ}}
\nc{\hcR}{\widehat{\cR}}
\nc{\hR}{\widehat{R}}
\nc{\hS}{\widehat{S}}
\nc{\hcS}{\widehat{\cS}}
\nc{\hT}{\widehat{T}}
\nc{\hU}{\widehat{U}}
\nc{\hV}{\widehat V}
\nc{\hcV}{\widehat \cV}
\nc{\hX}{\widehat X}
\nc{\hcZ}{\widehat \cZ}
\nc{\hcZbar}{\ol{\widehat \cZ}}

\nc{\heta}{\widehat{\eta}}
\nc{\hal}{\widehat \alpha}
\nc{\hbeta}{\widehat \beta}
\nc{\hphi}{\widehat{\phi}}
\nc{\hkap}{\hat{\kappa}}
\nc{\hchi}{\widehat{\chi}}
\nc{\hpsi}{\widehat{\psi}}
\nc{\hgam}{\widehat{\gam}}
\nc{\hPhi}{\hat{\Phi}}
\nc{\hPsi}{\hat{\Psi}}
\nc{\hGam}{\hat{\Gam}}
\nc{\omhat}{\widehat{\om}}
\nc{\htha}{\hat{\tha}}
\nc{\hrho}{\widehat{\rho}}
\nc{\hdel}{\widehat{\del}}

\nc{\w}{\wedge}

\nc{\vb}{\vec b}
\nc{\vc}{\vec c}
\nc{\vd}{\vec d}
\nc{\ve}{\vec e}
\nc{\vf}{\vec f}
\nc{\vg}{\vec g}
\nc{\vh}{\vec h}
\nc{\vp}{\vec p}
\nc{\vq}{\vec q}
\nc{\vr}{\vec r}
\nc{\vs}{\vec s}
\nc{\vv}{\vec v}
\nc{\vw}{\vec w}
\nc{\vx}{\vec x}
\nc{\vy}{\vec y}
\nc{\vz}{\vec z}

\nc{\vB}{\vec B}
\nc{\vC}{\vec C}
\nc{\vD}{\vec D}
\nc{\vE}{\vec E}
\nc{\vF}{\vec F}
\nc{\vG}{\vec G}
\nc{\vH}{\vec H}
\nc{\vP}{\vec P}
\nc{\vQ}{\vec Q}
\nc{\vR}{\vec R}
\nc{\vS}{\vec S}
\nc{\vV}{\vec V}
\nc{\vW}{\vec W}
\nc{\vX}{\vec X}
\nc{\vY}{\vec Y}
\nc{\vZ}{\vec Z}

\nc{\ol}{\overline}
\nc{\abar}{\ol{a}}
\nc{\bbar}{\ol{b}}
\nc{\cbar}{\ol{c}}
\nc{\dbar}{\ol{d}}
\nc{\ebar}{\ol{e}}
\nc{\fbar}{\ol{f}}
\nc{\gbar}{\ol{g}}
\nc{\ibar}{\ol{\imath}}
\nc{\jbar}{\ol{\jmath}}
\nc{\kbar}{\ol{k}}
\nc{\lbar}{\ol{l}}
\nc{\mbar}{\ol{m}}
\nc{\nbar}{\ol{n}}
\nc{\pbar}{\ol{p}}
\nc{\qbar}{\ol{q}}
\nc{\rbar}{\ol{r}}
\nc{\sbar}{\ol{s}}
\nc{\ubar}{\ol{u}}
\nc{\vbar}{\ol{v}}
\nc{\wbar}{\ol{w}}
\nc{\xbar}{\ol{x}}
\nc{\ybar}{\ol{y}}
\nc{\zbar}{\ol{z}}

\nc{\Abar}{\ol{A}}
\nc{\Bbar}{\ol{B}}
\nc{\cBbar}{\ol{\cB}}
\nc{\Cbar}{\ol{C}}
\nc{\Dbar}{\ol{D}}
\nc{\Ebar}{\ol{E}}
\nc{\Fbar}{\ol{F}}
\nc{\Gbar}{\ol{G}}
\nc{\Jbar}{\ol{J}}
\nc{\Kbar}{\ol{K}}
\nc{\cKbar}{\ol{\cK}}
\nc{\Lbar}{\ol{L}}
\nc{\cLbar}{\ol{\cL}}
\nc{\Mbar}{\ol{M}}
\nc{\Nbar}{\ol{N}}
\nc{\Pbar}{\ol{P}}
\nc{\Qbar}{\ol{Q}}
\nc{\Rbar}{\ol{R}}
\nc{\Sbar}{\ol{S}}
\nc{\Tbar}{\ol{T}}
\nc{\Ubar}{\ol{U}}
\nc{\Vbar}{\ol{V}}
\nc{\cVbar}{\ol{\cV}}
\nc{\Wbar}{\ol{W}}
\nc{\cWbar}{\ol{\cW}}
\nc{\Xbar}{{\overline X}}
\nc{\Ybar}{{\overline Y}}
\nc{\Zbar}{{\overline Z}}
\nc{\cZbar}{{\overline \cZ}}

\nc{\epsbar}{\ol{\epsilon}}
\nc{\albar}{\ol{\al}}
\nc{\Albar}{\ol{\Al}}
\nc{\betabar}{\ol{\beta}}
\nc{\Betabar}{\ol{\Beta}}
\nc{\lambar}{\ol{\lambda}}
\nc{\kapbar}{\ol{\kappa}}
\nc{\zetabar}{\ol{\zeta}}
\nc{\Zetabar}{\ol{\Zeta}}
\nc{\taubar}{\ol{\tau}}
\nc{\Taubar}{\ol{\Tau}}
\nc{\psibar}{\ol{\psi}}
\nc{\Psibar}{\ol{\Psi}}
\nc{\tpsibar}{\ol{\tpsi}}
\nc{\tPsibar}{\ol{\tPsi}}
\nc{\phibar}{\ol{\phi}}
\nc{\Phibar}{\ol{\Phi}}
\nc{\chibar}{\ol{\chi}}
\nc{\mubar}{\ol{\mu}}
\nc{\nubar}{\ol{\nu}}
\nc{\rhobar}{\ol{\rho}}
\nc{\ombar}{\ol{\om}}
\nc{\Ombar}{\ol{\Om}}
\nc{\Deltabar}{\ol{\Delta}}
\nc{\Thetabar}{\ol{\Theta}}
\nc{\xibar}{\ol{\xi}}
\nc{\Xibar}{\ol{\Xi}}

\nc{\Dthbar}{\ol{\rm D3}}

\nc{\fdot}{\dot{f}}
\nc{\gdot}{\dot{g}}
\nc{\pdot}{\dot{p}}
\nc{\qdot}{\dot{q}}
\nc{\rdot}{\dot{r}}
\nc{\sdot}{\dot{s}}
\nc{\tdot}{\dot{t}}
\nc{\udot}{\dot{u}}
\nc{\vdot}{\dot{v}}
\nc{\wdot}{\dot{w}}
\nc{\xdot}{\dot{x}}
\nc{\xddot}{\ddot{x}}
\nc{\ydot}{\dot{y}}
\nc{\zdot}{\dot{z}}
\nc{\yddot}{\ddot{y}}

\nc{\Adot}{\dot{A}}
\nc{\Bdot}{\dot{B}}
\nc{\Cdot}{\dot{C}}
\nc{\Udot}{\dot{U}}
\nc{\Vdot}{\dot{V}}
\nc{\Wdot}{\dot{W}}

\nc{\taudot}{\dot{\tau}}
\nc{\phidot}{\dot{\phi}}
\nc{\psidot}{\dot{\psi}}
\nc{\chidot}{\dot{\chi}}
\nc{\sinp}{s_{\phi}}
\nc{\cosp}{c_{\phi}}
\nc{\tanp}{t_{\phi}}
\nc{\spone}{s_{\phi_1}}
\nc{\cpone}{c_{\phi_1}}
\nc{\tpone}{t_{\phi_1}}
\nc{\sptwo}{s_{\phi_2}}
\nc{\cptwo}{c_{\phi_2}}
\nc{\tptwo}{t_{\phi_2}}
\nc{\spth}{s_{\phi_3}}
\nc{\cpth}{c_{\phi_3}}
\nc{\tpth}{t_{\phi_3}}
\nc{\calp}{c_{\al}}
\nc{\salp}{s_{\al}}

\nc{\csch}{{\rm csch}}
\nc{\sech}{{\rm sech}}

\nc{\cothzlami}{\coth(z-\lam_i)}
\nc{\coshzlami}{\cosh(z-\lam_i)}
\nc{\sinhzlami}{\sinh(z-\lam_i)}

\nc{\cothzlamj}{\coth(z-\lam_j)}
\nc{\coshzlamj}{\cosh(z-\lam_j)}
\nc{\sinhzlamj}{\sinh(z-\lam_j)}

\nc{\cothlamij}{\coth(\lam_i-\lam_j)}
\nc{\coshlamij}{\cosh(\lam_i-\lam_j)}
\nc{\sinhlamij}{\sinh(\lam_i-\lam_j)}

\nc{\bah}{{\mathbf {\hat{A}}}}
\nc{\bX}{{\mathbf X}}
\nc{\ba}{{\bf a}}
\nc{\bb}{{\bf b}}
\nc{\bc}{{\bf c}}
\nc{\bd}{{\bf d}}
\nc{\bg}{{\bf g}}
\nc{\bk}{{\bf k}}
\nc{\bl}{{\bf l}}
\nc{\bm}{{\bf m}}
\nc{\bn}{{\bf n}}
\nc{\bo}{{\bf o}}
\nc{\bp}{{\bf p}}
\nc{\bq}{{\bf q}}
\nc{\br}{{\bf r}}
\nc{\bs}{{\bf s}}
\nc{\bt}{{\bf t}}
\nc{\bu}{{\bf u}}
\nc{\bv}{{\bf v}}
\nc{\bw}{{\bf w}}
\nc{\bx}{{\bf x}}
\nc{\by}{{\bf y}}
\nc{\bz}{{\bf z}}
\nc{\bom}{{\bf \om}}
\nc{\bombar}{{\mathbf \ombar}}
\nc{\bPhi}{{\bf \Phi}}

\nc{\rma}{{\rm a}}
\nc{\rmb}{{\rm b}}
\nc{\rmc}{{\rm c}}
\nc{\rmd}{{\rm d}}
\nc{\rmg}{{\rm g}}
\nc{\rk}{{\rm k}}
\nc{\rml}{{\rm l}}
\nc{\rmm}{{\rm m}}
\nc{\rmn}{{\rm n}}
\nc{\rmo}{{\rm o}}
\nc{\rmp}{{\rm p}}
\nc{\rmq}{{\rm q}}
\nc{\rmr}{{\rm r}}
\nc{\rms}{{\rm s}}
\nc{\rmt}{{\rm t}}
\nc{\rmu}{{\rm u}}
\nc{\rmv}{{\rm v}}
\nc{\rmw}{{\rm w}}
\nc{\rmx}{{\rm x}}
\nc{\rmy}{{\rm y}}
\nc{\rmz}{{\rm z}}

\nc{\dal}{\dot{\al}}
\nc{\thadot}{\dot{\tha}}
\nc{\thab}{\bar{\theta}}
\nc{\thal}{\theta^{\al}}
\nc{\thdal}{\bar{\theta}^{\dal}}

\nc{\thsigthm}{\tha \sigma^m \thab}
\nc{\thsigthn}{\tha \sigma^n \thab}

\nc{\Dal}{D_{\al}}
\nc{\Ddal}{\bar{D}_{\dal}}
\nc{\CDal}{{\cal D}_{\al}}
\nc{\CDdal}{\bar{\cal D}_{\dal}}

\nc{\eq}[1]{{(\ref{#1})}}
\nc{\eqtwo}[2]{{(\ref{#1},\ref{#2})}}
\nc{\eqthree}[3]{(\ref{#1},\ref{#2},\ref{#3})}
\nc{\eqfour}[4]{(\ref{#1},\ref{#2},\ref{#3},\ref{#4})}
\nc{\eqfive}[5]{(\ref{#1},\ref{#2},\ref{#3},\ref{#4,\ref{#5}})}
\nc{\non}{\nonumber}
\nc{\Fzero}{F_{(0)}}
\nc{\Ftwo}{F_{(2)}}
\nc{\Ffour}{F_{(4)}}
\nc{\Fone}{F_{(1)}}
\nc{\Fthree}{F_{(3)}}
\nc{\Ffive}{F_{(5)}}
\nc{\Fn}{F_{(n)}}
\nc{\Fp}{F_{(p)}}

\nc{\tFzero}{\tF_{(0)}}
\nc{\tFtwo}{\tF_{(2)}}
\nc{\tFfour}{\tF_{(4)}}
\nc{\tFone}{\tF_{(1)}}
\nc{\tFthree}{\tF_{(3)}}
\nc{\tFfive}{\tF_{(5)}}
\nc{\tFn}{\tF_{(n)}}
\nc{\tFp}{\tF_{(p)}}

\nc{\Czero}{C_{(0)}}
\nc{\Ctwo}{C_{(2)}}
\nc{\Cfour}{C_{(4)}}
\nc{\Cone}{C_{(1)}}
\nc{\Cthree}{C_{(3)}}
\nc{\Cfive}{C_{(5)}}
\nc{\Cn}{C_{(n)}}

\nc{\equ}{{\rm eq}}
\def\Im{{\rm Im \hspace{0.5mm} }}

\def\Re{{\rm Re \hspace{0.5mm}}}

\nc{\vol}{{\rm vol}}
\nc{\Ainf}{A_{\infty}}
\nc{\End}{{\rm End}}
\nc{\Ext}{{\rm Ext}}
\nc{\IIB}{{\rm IIB}}
\nc{\Ad}{{\rm Ad}}
\nc{\IIA}{{\rm IIA}}
\nc{\AdS}{{\rm AdS}}
\nc{\CFT}{{\rm CFT}}
\nc{\diag}{{\rm diag}}
\nc{\Log}{{\rm Log}}
\nc{\Dslash}{\ensuremath \raisebox{0.025cm}{\slash}\hspace{-0.32cm} D}
\nc{\cDslash}{\ensuremath \raisebox{0.025cm}{\slash}\hspace{-0.32cm} \cD}
\nc{\omslash}{\om\!\!\!/}
\nc{\no}{\!:\!\!}
\nc{\ointdz}{\oint\frac{dz}{2\pi i}}
\nc{\ointdzone}{\oint\frac{dz_1}{2\pi i}}
\nc{\ointdztwo}{\oint\frac{dz_2}{2\pi i}}
\nc{\ointdzb}{\oint\frac{d\zbar}{2\pi i}}
\nc{\ointdzbone}{\oint\frac{d\zbar_1}{2\pi i}}
\nc{\ointdzbtwo}{\oint\frac{d\zbar_2}{2\pi i}}
\nc{\dz}{\frac{dz}{2\pi i}}
\nc{\dzb}{\frac{d\zbar}{2\pi i}}
\nc{\bpm}{\begin{pmatrix}}
\nc{\epm}{\end{pmatrix}}
 \nc{\bitem}{\begin{itemize}}
 \nc{\eitem}{\end{itemize}}
 \nc{\exercise}{\vskip 2mm \noindent {\bf Exercise:}}
 \nc{\definition}{\vskip 2mm \noindent {\bf Definition:}}
 
\begin{document}
%%%%%%%%%%%%%%%%%%%%%%%%%%%%%%%%%%%%%%%%%%%%%%%%%%%%%%%%%%%%%%%%%
%%%%%%%%%%%%%%%%%%%%%%%%%%%%%%%%%%%%%%%%%%%%%

\begin{center}
\baselineskip=13pt {\LARGE \bf{Static BPS Black Holes in AdS$_4$ \\
with General Dyonic Charges}}
\vskip1.5cm
Nick Halmagyi\\ 
\vskip0.5cm
\textit{Sorbonne Universit\'es, UPMC Paris 06,  \\ 
UMR 7589, LPTHE, 75005, Paris, France \\
and \\
CNRS, UMR 7589, LPTHE, 75005, Paris, France}\\
\vskip1cm
halmagyi@lpthe.jussieu.fr \\ 
\vskip1cm

\end{center}

\begin{abstract}
We complete the study of static BPS, asymptotically AdS$_4$ black holes within $\cN=2$ FI-gauged supergravity and where the scalar manifold is a homogeneous very special K\"ahler manifold. We find the analytic form for the general solution to the BPS equations, the horizon appears as a double root of a particular quartic polynomial whereas in previous work this quartic polynomial further factored into a pair of double roots. A new and distinguishing feature of our solutions is that the phase of the supersymmetry parameter varies throughout the black hole. The general solution has $2n_v$ independent parameters; there are two algebraic constraints on $2n_v+2$ charges, matching our previous analysis on BPS solutions of the form AdS$_2\times \Sig_g$. As a consequence we have proved that every BPS geometry of this form can arise as the horizon geometry of a BPS AdS$_4$ black hole. When specialized to the STU-model our solutions uplift to M-theory and describe a stack of M2-branes wrapped on a Riemman surface in a Calabi-Yau fivefold with internal angular momentum.
\end{abstract}

\newpage 
%%%%%%%%%%%%%%%%%%%%%%%%%%%%%%%%%%%%
\section{Introduction}
%%%%%%%%%%%%%%%%%%%%%%%%%%%%%%%%%%%%

The study of BPS black holes in flat space has provided important insights into the nature of quantum gravity \cite{Strominger:1996sh} and it is natural to search for a similar understanding of BPS black holes in asymptotically AdS space. Indeed there is hope that one might ultimately be able to use the holographic duality of the background AdS space \cite{Maldacena:1997re, Gubser:1998bc, Witten:1998qj} to understand the dynamics of black holes and these holographic dualities are certainly under the best control for supersymmetric theories. 

Whereas the complete solution for asymptotically flat BPS black holes in four dimensional $\cN=2$ ungauged supergravity was found some time ago \cite{Behrndt:1997ny, Bates:2003vx} it is only somewhat recently that there has been progress in charting out the solution space of asymptotically AdS$_4$ static BPS black holes in gauged $\cN=2$ supergravity. In this work we complete the study of the solution space of static BPS black holes in $\cN=2$ FI-gauged supergravity\footnote{FI-gauged supergravity refers to the Fayet-Iliopoulos gauging, namely where the $U(1)_R\subset SU(2)_R$ is gauged}. To be more precise we require that the vector-multiplet scalar manifold $\cM_v$ is homogenous and very special K\"ahler or in other words that it is a coset and that there is a symplectic duality frame where the prepotential is cubic.

A notable development in this field was the solution of Cacciatori-Klemm \cite{Cacciatori:2009iz} who studied a restricted set of background charges in the STU-model and found a class of solutions depending on four charges with one constraint, leaving a three-dimensional solution space\footnote{This solution was further studied in \cite{Dall'Agata:2010gj, Hristov:2010ri}}. This constraint is a BPS refinement of the Dirac quantization condition due to the charged gravitini and is of course absent in ungauged supergravity and the study of asymptotically flat black holes. 

We have recently considered \cite{Halmagyi:2013qoa} these general $\cN=2$ FI-gauged supergravity theories and exactly solved the BPS conditions for geometries of the form AdS$_2\times \Sig_g$ where $\Sig_g\in \{S^2,\RR^2,\HH^2\}$ is (a covering space of) a Riemann surface of genus $g$. We found that the entropy of the solution is related to the famous quartic invariant of special geometry and indeed this invariant will prove to be indispensable in our current analysis as well. An interesting result from the analysis of horizon geometries was that the solution space is\footnote{$n_v$ is the number of vector multiplets} $2n_v$-dimensional; there are $n_v+1$ magnetic charges $p^\Lam$, along with $n_v+1$ electric charges $q_\Lam$ and there are two constraints from the BPS conditions. One constraint is the Dirac quantization condition mentioned earlier but in addition there is another constraint given by \eq{constraint} below. Consequently it was conjectured that the solution space of BPS black holes is also $2n_v$-dimensional. In this work we prove this conjecture by constructing these solutions explicitly.

In previous work we have generalized the solution of \cite{Cacciatori:2009iz} in two ways; for theories with a homogeneous very special K\"ahler scalar manifold we presented axion-free solutions which depend on $n_v$ independent charges and give the CK solution when restricted to the STU model. In addition, in \cite{Halmagyi:2013uza} we showed how to use unbroken duality symmetries to generate two new charges in the STU-model. For the STU-model in particular, this gives a five-dimensional solution space whereas the analysis of \cite{Halmagyi:2013qoa} predicts that there should be a six-dimensional solution space. Recently it was shown \cite{Katmadas:2014faa} that this family of solutions with non-trivial axions can be generalized from the STU-model to theories where $\cM_v$ is a homogeneous very special K\"ahler manifold but still, the dimension of the solution space was $2n_v-1$, one less than conjectured in \cite{Halmagyi:2013qoa}. The final dimension of the solution space indeed exists as we constructively prove in the current work.

This paper is organized as follows. In section \ref{sec:Equations} we review the BPS equations and outline our ansatz. In section \ref{sec:Solution} we present our solution to the BPS equations. In appendix \ref{app:Special} we summarize various details of special geometry which are needed for the current work.

%%%%%%%%%%%%%%%%%%%%%%%%%%%%%%%%%%%%
\section{The Equations and the Ansatz} \label{sec:Equations}
%%%%%%%%%%%%%%%%%%%%%%%%%%%%%%%%%%%%

The standard ansatz for static AdS$_4$ black holes is
\be
ds_4^2 = -e^{2U}dt_2 + e^{-2U} dr^2 + e^{2(V-U)} d\Sig_g^2
\ee
where $d\Sig_g^2$ is the uniform metric on $(S^2,T^2,\HH^2)$ with corresponding curvature $\kappa=(1,0,-1)$. The scalar fields are radially dependent $z^i=z^i(r)$ and the gauge fields just contribute through the conserved charges
\bea
p^\Lam = \frac{1}{4\pi }\int_{\Sig_g} F^\Lam\,,\quad\quad
q_\Lam = \frac{1}{4\pi }\int_{\Sig_g} G^\Lam
\eea
where the dual field strength is given by
\be
G_\Lam = \cR_{\Lam \Sig} F^\Sig -\cI_{\Lam \Sig} *_4 F^\Sig\,.
\ee
The charges and gauging parameters\footnote{The gravitini are charged with respect to the $U(1)$ gauge fields $A_\mu^\Lam$ by $\cG$, it is always possible to find a symplectic frame where these couplings are purely electric and so the gravitini are just minimally coupled to the $A_\mu^\Lam$} are naturally assembled into symplectic vectors
\be
\cQ=\bpm p^\Lam \\ q_\Lam\epm\,,\quad\quad \cG=\bpm g^\Lam \\ g_\Lam \epm\,.
\ee
A static BPS AdS$_4$ black hole amounts to a domain wall interpolating between AdS$_4$ in the UV to AdS$_2\times \Sig_g$ in the IR, the effective cosmological constant varies along this flow due to the non-trivial profiles for scalar fields which contribute to the cosmological constant through the scalar potential.

For a particular $\cG$ and $\cQ$ in the STU-model\footnote{In the symplectic frame where the gauging parameters are purely electric, the charges studied in \cite{Cacciatori:2009iz} were purely magnetic}, in \cite{Cacciatori:2009iz} the equations for $\frac{1}{4}$-BPS black holes were found to simplify considerably and the exact solution was found. These solutions can be lifted in M-theory and interpreted as wrapped M2-branes along the lines of the work of Maldacena-Nunez \cite{Maldacena:2000mw}. Indeed restricted classes of such wrapped M2-branes had already been considered numerically in \cite{Gauntlett:2001qs} but the solution of \cite{Cacciatori:2009iz} was the first exact supergravity solution\footnote{We are excluding the elementary solutions with constant scalar fields which for four dimensional black holes were considered in \cite{Caldarelli1999}} in any dimension of a stack of wrapped branes which has an AdS factor in the IR. 

In four dimensions one has available the tools of special geometry and in \cite{Dall'Agata:2010gj} a duality covariant form of the BPS equations was derived\footnote{We use the standard notation for the sections 
\be
\cV=\bpm L^\Lam \\ M_\Lam \epm \ =\ e^{K/2} \bpm X^\Lam \\ F_\Lam  \epm
\ee
}:
\bea
2 e^{2V} \del_r \Bslb \Im\blp e^{-i\psi}e^{-U}\cV \brp \Bsrb &=& 8 e^{2(V-U)} \Re\blp e^{-i\psi} \cL\brp \Re \blp e^{-i\psi}\cV \brp -\cQ -e^{2(V-U)} \Om \cM \cG \label{DGEq1}\\
\del_r \blp e^V \brp &=& 2 e^{V-U} \Im \blp  e^{-i\psi} \cL\brp \label{DGEq2} \\
\psi'+A_r&=& - e^{U-2V} \Re (e^{-i\psi} \cZ) - e^{-U}\Im (e^{-i\psi} \cL) \label{DGEq3}
\eea
where $\psi$ is the phase of the supersymmetry parameter and we have defined the standard symplectic scalars
\be
\cL=\langle \cG,\cV\rangle \,, \quad\quad \cZ=\langle \cQ,\cV\rangle\,.
\ee
In addition one must impose the Dirac quantization condition specialized for these BPS equations:
\be
\langle \cG, \cQ \rangle =-\kappa\,. \label{DiracBPS}
\ee
The equation  \eq{DGEq1} is a real symplectic vector with $2n_v+2$ components which can be extracted by contracting \eq{DGEq1} with $\{\cV,\cVbar,D_i\cV,D_{\ibar} \cVbar \}$. In this way, one obtains differential equations for the phase $\psi$, the metric mode $e^U$ and the $n_v$ complex scalars $z^i$. Comparing the resulting equation for $\psi$ with \eq{DGEq3} one obtains the algebraic constraint
\be
 \Re\blp e^{-i\psi} \cL\brp  =e^{2(V-U)}  \Im\blp e^{-i\psi} \cZ\brp  \label{DGEq4}
\ee
 which can then be enforced at the expense of \eq{DGEq3}. To be clear, an equivalent complete set of BPS conditions is \eq{DGEq1},\eq{DGEq2},\eq{DiracBPS},\eq{DGEq4} and this is the set we will solve in the next section. It may be interesting to note that the constraint \eq{DGEq4} is a generalization of the constraint $0=\Im\blp e^{-i\psi} \cZ\brp$  found for single-center half-BPS black holes in ungauged supergravity \cite{Denef:2000nb}.

Using these equations, the analytic solution of \cite{Cacciatori:2009iz} was generalized to theories where $\cM_v$ is a homogeneous very special K\"ahler manifold in \cite{Gnecchi:2013mta}. In these examples, the axions vanish identically and the complex symplectic sections become either real or imaginary leading to a sharp simplification of the equations \eq{DGEq1}-\eq{DGEq2} and in these example both sides of \eq{DGEq4} vanish identically. The solutions of \cite{Gnecchi:2013mta} are parameterized by $n_v$ charges out of the $2n_v+2$ components of $\cQ$ and in \cite{Halmagyi:2013uza} it was observed that one can use unbroken duality symmetries to generate axions.  

The BPS equation \eq{DGEq1} was recently refined in \cite{Katmadas:2014faa} using the quartic invariant $I_4$ and its derivative\footnote{to be precise this rewriting is only valid when $\cM_v$ is a homogeneous very special K\"ahler manifold, we have included substantial background about the quartic invariant $I_4$ and its derivative $I_4'$ in appendix \ref{app:quartic}.}, which further simplifies the analysis of the models with general gauging parameters $\cG$ and charges $\cQ$. Specifically, using the identity \eq{I4pQImVImV} one immediately finds that \eq{DGEq1},\eq{DGEq2} and \eq{DGEq4} become
\bea
0&=& 2 e^V \del_r(\Im \tcV) - I'_4 \blp \Im \tcV, \Im \tcV,\cG\brp+\cQ \label{BPSV1}\\
(e^V)'&=& 2 \langle \cG, \Im \tcV \rangle  \label{BPSV2} \\
 \langle \cG,\Re \tcV \rangle   &=& e^{2(V-U)}  \langle \cQ,\Im \tcV \rangle  \label{BPSV3}
\eea
where
\be
\tcV=e^{V-U} e^{-i\psi} \cV\,.
\ee
The key step in arriving at \eq{BPSV1} is to remove the pesky $\Om\cM\cG$ and $\Re\cV$ terms in \eq{DGEq1} leaving an equation purely in terms of $\Im\tcV$ and $e^V$. The main result of the current work is to solve \eq{BPSV1}-\eq{BPSV3} in complete generality subject to the boundary conditions of AdS$_4$ in the UV and AdS$_2\times \Sig_g$ in the IR.

The key insight of Cacciatori and Klemm in \cite{Cacciatori:2009iz} was the ansatz for $e^V$:
\be\label{VCK}
e^V|_{CK}=\frac{r^2}{R}-v_0\,,
\ee
where $R$ is the radius of AdS$_4$ and $v_0>0$. In \eq{VCK} $e^V$ is completely fixed by the UV and IR boundary conditions whereas in principle $e^V$ could be a power series in $\frac{1}{r}$ and indeed that is what we will find\footnote{This contradicts the analysis in \cite{Katmadas:2014faa} where such solutions were excluded}. In this work we generalize \eq{VCK} to the following ansatz\footnote{we have shifted the radial co-ordinate to move the horizon to $r=0$}
\be\label{eVansatz}
e^V= r\sqrt{v_4 r^2 + v_3 r + v_2}\,.
\ee
Clearly $e^{2V}$ has a double root at $r=0$, a familiar feature of charged extremal black holes namely that the two horizons coincide, the remaining two roots however need not coincide. 
The ansatz for $\Im \tcV$ requires further inspiration, we find that the following is ultimately justified by our success:
\be \label{tcVansatz}
\Im \tcV = e^{-V} \Bslb A_1 r + A_2 r^2 + A_3 r^3 \Bsrb\,,
\ee
where $A_i$ are the  symplectic vectors which will be explicitly solved for.

%%%%%%%%%%%%%%%%%%%%%%%%%%%%%%%%%%%%%%%%%%%%%
\section{The Solution}\label{sec:Solution}
%%%%%%%%%%%%%%%%%%%%%%%%%%%%%%%%%%%%%%%%%%%%%
%%%%%%%%%%%%%%%%%%%%%%%%%%%%%%%%%%%%%%%%%%%%%
\subsection{The General Solution}
%%%%%%%%%%%%%%%%%%%%%%%%%%%%%%%%%%%%%%%%%%%%%
When evaluated on the ansatz \eq{eVansatz} and \eq{tcVansatz} and expanded order by order, the BPS equation \eq{BPSV1} yields
\bea
I'_4(A_3,A_3,\cG)&=& 2 \langle \cG,A_3 \rangle A_3 \label{BPSorder1}  \\
I'_4(A_2,A_3,\cG)&=& 2 \langle \cG,A_2 \rangle A_3 \label{BPSorder2}  \\
2I'_4(A_1,A_3,\cG)+I'_4(A_2,A_2,\cG)&=& \langle \cG,A_3 \rangle (\cQ-2A_1)  +8 \langle \cG,A_1 \rangle A_3+\frac{4}{3}\langle \cG,A_2 \rangle A_2\label{BPSorder3}  \\
3I'_4(A_1,A_2,\cG)&=& 6\langle \cG,A_1 \rangle A_2 + 2\langle \cG,A_2 \rangle (\cQ-A_1) \label{BPSorder4} \\
I'_4(A_1,A_1,\cG)&=& 2 \langle \cG,A_1 \rangle \cQ\label{BPSorder5} 
\eea
where we have used the solution to \eq{BPSV2}
\bea
v_{i+1}&=& \frac{4}{i+1}\langle \cG,A_i \rangle\,.
\eea
The constraint \eq{BPSV3} expanded order by order in $r$ gives
\bea
\langle A_3,A_2\rangle&=& 0 \label{Constrorder1} \\
 2 \langle A_1,A_3\rangle&=&  \langle \cQ,A_3\rangle  \label{Constrorder2} \\ 
 \langle A_1,A_2\rangle&=&  \langle \cQ,A_2\rangle \label{Constrorder3}  \\
 0&=&   \langle \cQ,A_1\rangle  \label{Constrorder4}\,. 
\eea
It should be noted that this whole system of equations is vastly overdetermined and for a solution to exist the system must be highly redundant. One starts solving this system by first re-analyzing the UV \eq{BPSorder1} and IR \eq{BPSorder5},\eq{Constrorder4} but the intermediate equations involve $A_2$ in nontrivial ways.

The UV boundary conditions for AdS$_4$ provide $A_3$ as the solution to \eq{BPSorder1} and one finds
\bea
A_3&=& \frac{6I'_4(\cG)}{\sqrt{I_4(\cG)}} \,,\quad\quad v_4= \frac{1}{R_{AdS_4}^2}=\sqrt{I_4(\cG)}
\eea
where we have appealed to the analysis of \cite{Gnecchi:2013mta} to fix the overall normalization.

The BPS equations are non-linear in terms of the $A_i$ and are somewhat challenging to solve directly. Making use of the various relations contained in appendix \ref{app:quartic} we have found the following explicit solution:
\bea
A_1&=&a_{1} I'_4(\cG)+a_{2} I'_4(\cG,\cG,\cQ)+a_{3} I'_4(\cG,\cQ,\cQ)+a_{4} I'_4(\cQ)\,,
\eea
where the constants $a_{j}$ are given by 
\bea
a_{1}&=&\frac{I_4(\cQ) I_4(\cG,\cG,\cG,\cQ)^2-I_4(\cG)I_4(\cG,\cQ,\cQ,\cQ)^2  }{ I_4(\cG,\cG,\cG,\cQ)^2 \blp \kappa\, I_4(\cG,\cG,\cQ,\cQ) + 2  \langle I'_4(\cG),I'_4(\cQ)\rangle\brp } \label{a1Sol}\\
a_{2}&=&-\frac{1}{12} \frac{I_4(\cG,\cQ,\cQ,\cQ)  }{ \kappa\, I_4(\cG,\cG,\cQ,\cQ) + 2  \langle I'_4(\cG),I'_4(\cQ)\rangle }\\
a_{3}&=&\frac{I_4(\cG)I_4(\cG,\cQ,\cQ,\cQ)^2 -I_4(\cQ) I_4(\cG,\cG,\cG,\cQ)^2 }{ 2 I_4(\cG,\cG,\cG,\cQ) I_4(\cG,\cQ,\cQ,\cQ) \blp \kappa\, I_4(\cG,\cG,\cQ,\cQ) + 2  \langle I'_4(\cG),I'_4(\cQ)\rangle\brp }  \\
a_{4}&=& \frac{1}{6} \,\frac{I_4(\cG,\cG,\cG,\cQ)}{ \kappa\, I_4(\cG,\cG,\cQ,\cQ) + 2  \langle I'_4(\cG),I'_4(\cQ)\rangle }\,. \label{a4Sol}
\eea

Our solution for $A_2$ is somewhat more complicated, the ansatz is the same as for $A_1$
\bea
A_2&=&b_! I'_4(\cG)+b_2 I'_4(\cG,\cG,\cQ)+b_3 I'_4(\cG,\cQ,\cQ)+b_4 I'_4(\cQ)
\eea
and we find that the constants $b_{j}$ are given by 
\bea
b_1&=&\frac{2b_3}{}\Bslb  \frac{I_4(\cQ) I_4(\cG,\cG,\cG,\cQ)}{I_4(\cG) I_4(\cG,\cQ,\cQ,\cQ)} - \frac{2I_4(\cG,\cQ,\cQ,\cQ)}{I_4(\cG,\cG,\cG,\cQ)} + \frac{\kappa}{18} \frac{I_4(\cG,\cG,\cG,\cQ)I_4(\cG,\cQ,\cQ,\cQ)^2}{ I_4(\cG)\Pi_3} \Bsrb \non \\
&&\label{b1Sol} \\
b_2&=&-\frac{b_3}{6} \frac{I_4(\cG,\cG,\cG,\cQ)}{I_4(\cG)}\, \frac{\Pi_1}{ \Pi_3}   \\
&& \non \\
b_3^2 &=&-\frac{9}{16}  I_4(\cG)^{3/2} I_4(\cG,\cG,\cG,\cQ) I_4(\cG,\cQ,\cQ,\cQ) \Pi_3^2  .\bslb 2\kappa I_4(\cQ) I_4(\cG,\cG,\cG,\cQ)^2 +  I_4(\cG,\cQ,\cQ,\cQ) \Pi_2 \bsrb^{-1} \non \\
&&\quad\quad  \bslb-\kappa I_4(\cG,\cG,\cG,\cQ)^2 I_4(\cG,\cQ,\cQ,\cQ)  \Pi_1  +18 I_4(\cG) \Pi_3^2 \bsrb^{-1}  \\
&& \non \\
b_4 &=&- \frac{2b_3\kappa }{3} \frac{I_{4}(\cG,\cG,\cG,\cQ)I_4(\cG,\cQ,\cQ,\cQ)}{\Pi_3 } \label{b4Sol}
\eea
where we have defined
\bea
\Pi_1&=& 2I_4(\cQ) I_4(\cG,\cG,\cG,\cQ) \kappa +I_4(\cG,\cQ,\cQ,\cQ) \langle I'_4(\cG),I'_4(\cQ) \rangle \non \\
\Pi_2&=& 2I_4(\cQ) I_4(\cG,\cQ,\cQ,\cQ) \kappa +I_4(\cG,\cG,\cG,\cQ) \langle I'_4(\cG),I'_4(\cQ) \rangle \\
\Pi_3&=& 4I_4(\cQ) I_4(\cG,\cG,\cG,\cQ)\kappa + I_4(\cG,\cQ,\cQ,\cQ)\langle I'_4(\cG),I'_4(\cQ) \rangle\,. \non 
\eea
Note that the BPS equations have a symmetry $A_2\ra -A_2$ and as such the overall sign of $b_3$ in our expressions is undetermined by the BPS conditions and must be fixed by demanding regularity of the solution. A priori one might have expected that the ansatz for $A_i$ involved terms proportional to the symplectic vectors $\cG$ and $\cQ$, we have considered such terms and found their coefficients to vanish.

With these expressions for $A_i$ we get the following result for the expansion of $e^V$:
\bea
v_2&=&-\frac{ I_4(\cG,\cG,\cG,\cQ)^3 I_4(\cG,\cQ,\cQ,\cQ)^3+36 \bslb I_4(\cQ) I_4(\cG,\cG,\cG,\cQ)^2- I_4(\cG) I_4(\cG,\cQ,\cQ,\cQ)^2\bsrb^2}{9 I_4(\cG,\cG,\cG,\cQ)^2 I_4(\cG,\cQ,\cQ,\cQ)^2 \blp \kappa I_4(\cG,\cG,\cQ,\cQ) +2 \langle I'_4(\cG),I'_4(\cQ) \rangle \brp} \non \\
&& \\
v_3&=& b_3\Bslb 3 I_4(\cG,\cG,\cQ,\cQ) -\frac{I_4(\cG,\cG,\cG,\cQ)^2}{6I_4(\cG)}-\frac{24 I_4(\cG)I_4(\cG,\cQ,\cQ,\cQ)}{I_4(\cG,\cG,\cG,\cQ)}  \non \\
&& + \frac{\kappa I_4(\cG,\cG,\cG,\cQ)^2 I_4(\cG,\cG,\cQ,\cQ) I_4(\cG,\cQ,\cQ,\cQ)}{12 I_4(\cG) \Pi_3} \Bsrb 
\eea

In addition, the constraint \eq{BPSV3} in the IR \eq{Constrorder4} gives the constraint found in \cite{Halmagyi:2013qoa}:
\bea 
0&=& 4I_4(\cG) I_{4}(\cG,\cQ,\cQ,\cQ)^2+4I_{4}(\cQ) I_{4}(\cQ,\cG,\cG,\cG)^2 \non \\
&&-  I_4(\cG,\cQ,\cQ,\cQ) I_4(\cG,\cG,\cQ,\cQ)I_{4}(\cQ,\cG,\cG,\cG)   \label{constraint}\,.
\eea
Since this constraint is independent of the radius it must be imposed on the whole black hole. We find that with this constraint enforced, the above solution solves each of \eq{Constrorder1}-\eq{Constrorder4}. The expressions above for $\{a_i,b_j,v_k\}$ have been evaluated subject to \eq{constraint} although in principle one could have used \eq{constraint} to eliminate $I_4(\cG,\cG,\cQ,\cQ)$. We have used what we find to be a more concise presentation.

We emphasize that solving for $\Im\tcV$ constitutes a complete solution of the problem, from which the metric functions $(e^U,e^V)$, the scalar fields $z^i$ and the phase of the supersymmetry parameter $\psi$ can all be obtained as follows. Using the identity (which follows from \eq{I4pQImVImV} evaluated with $A\ra \Im \tcV$):
\be
I'_4(\Im\tcV,\Im\tcV,\Im\tcV)= 4 \Im\bslb \langle \Im\tcV,\tcV \rangle\bsrb \Im \tcV + 8 \Re\bslb \langle \Im\tcV,\tcV \rangle\bsrb \Re \tcV  - e^{2(V-U)}\Om \cM \Im\tcV 
\ee
we find
\be \label{RetcV}
\Re \tcV =-\frac{ I'_4(\Im\tcV)} {2\sqrt{\, I_4(\Im\tcV)}}\,.
\ee
Having the expression for the complex sections $\tcV$ it is straightforward to read off the complex scalar fields:
\be
z^i =\frac{\tcL^i}{\tcL^0}\,.
\ee
Contracting \eq{RetcV}  with $\Im\tcV$ we get the remaining metric function from
\be
e^{2(V-U)}=4 \sqrt{ I_4(\Im\tcV)}\,.
\ee
The phase of the supersymmetry parameter $\psi$ can be extracted from
\be
e^{2i\psi}= \frac{\tcLbar^0}{\tcL^0}
\ee
and is generically non-constant as opposed to all previous solutions in \cite{Cacciatori:2009iz, Gnecchi:2013mta, Halmagyi:2013uza, Katmadas:2014faa} where the supersymmetry phase is constant. Note that the numerical solution presented in \cite{Halmagyi:2013sla} for gauged supergravity with hypermultiplets also has a non-constant phase for the spinor.

While the explicit solution \eq{a1Sol}-\eq{a4Sol} and \eq{b1Sol}-\eq{b4Sol} is satisfying in the sense that it is exhaustive within the class of theories considered, its explicit form is not exactly edifying. The crucial step in finding these solutions was starting from the metric ansatz in \eq{eVansatz}, which for various reasons this was incorrectly excluded in \cite{Katmadas:2014faa}. It seems plausible that using the tools of special geometry one could find a a more concise representation of this solution, which may possibly be required for this solution to have applications to generalizations discussed in the introduction.

%%%%%%%%%%%%%%%%%%%%%%%%%%%%%%%%%%%%%%%%%%%%%
\subsection{The Limit $I_4(\cG,\cG,\cG,\cQ)=I_{4}(\cG,\cQ,\cQ,\cQ)=0$}
%%%%%%%%%%%%%%%%%%%%%%%%%%%%%%%%%%%%%%%%%%%%%
To observe the limit $I_4(\cG,\cG,\cG,\cQ)=I_{4}(\cG,\cQ,\cQ,\cQ)=0$, it is useful to note that the constraint \eq{constraint} implies
\be
\frac{I_{4}(\cG,\cQ,\cQ,\cQ)}{I_4(\cG,\cG,\cG,\cQ)} =\frac{I_4(\cG,\cG,\cQ,\cQ)\pm \sqrt{I_4(\cG,\cG,\cQ,\cQ)^2-64 I_4(\cG)I_4(\cQ)}}{8 I_4(\cG)}
\ee
which is finite in this limit. We find the following explicit expressions
\bea
a_1&=& \frac{a_3 \blp \sqrt{I_4(\cG,\cG,\cQ,\cQ)^2-64 I_4(\cQ) I_4(\cG)} -I_4(\cG,\cG,\cQ,\cQ) \brp}{24 I_4(\cG)} \\
a_2&=& 0 \\
a_3&=&- \frac{\sqrt{I_4(\cG,\cG,\cQ,\cQ)^2-64 I_4(\cQ) I_4(\cG)}}{8\blp \kappa I_4(\cG,\cG,\cQ,\cQ) + 2 \langle I_4(\cG),I_4(\cQ) \rangle \brp}\\
a_4&=&0
\eea
as well as
\bea
b_1&=& \frac{a_3 \blp 3 \sqrt{I_4(\cG,\cG,\cQ,\cQ)^2-64 I_4(\cQ) I_4(\cG)} -I_4(\cG,\cG,\cQ,\cQ) \brp}{24 I_4(\cG)} \\
b_2&=& 0 \\
b_3&=& \frac{ I_4(\cG)^{1/4}}{4\sqrt{-\kappa I_4(\cG,\cG,\cQ,\cQ)- 2 \langle I_4(\cG),I_4(\cQ) \rangle} } \\
b_4&=&0 
\eea
and
\bea
v_2&=& \frac{64 I_4(\cG) I_4(\cQ) - I_4(\cG,\cG,\cQ,\cQ)^2}{4\blp \kappa I_4(\cG,\cG,\cQ,\cQ)+ 2 \langle I_4(\cG),I_4(\cQ) \rangle   \brp} \\
v_3&=& \frac{ 3I_4(\cG)^{1/4} \sqrt{64 I_4(\cG) I_4(\cQ) - I_4(\cG,\cG,\cQ,\cQ)^2}}{4\sqrt{\kappa I_4(\cG,\cG,\cQ,\cQ)+ 2 \langle I_4(\cG),I_4(\cQ) \rangle} }
\eea
This limit corresponds to those solutions which can be obtained by a duality transformation from the solutions in \cite{Gnecchi:2013mta}, a distinguishing feature is that the phase of the supersymmetry parameter (which is duality invariant) is constant.

%%%%%%%%%%%%%%%%%%%%%%%%%%%%%%%%%%%%%%%%%%%%%
\section{Conclusions}
%%%%%%%%%%%%%%%%%%%%%%%%%%%%%%%%%%%%%%%%%%%%%

We have solved the general static BPS black hole in $\cN=2$ FI-gauged supergravity and found that the full solution depends on $2n_v$ out of $2n_v+2$ charges. The two constraints are given by the BPS Dirac quantization condition \eq{DiracBPS} as well as \eq{constraint}, where the latter has its origin in \eq{DGEq4}. When restricted to the STU-model, all these solutions lift to M-theory via an embedding \cite{Cvetic1999b} of the STU-model into the $\cN=8$ de-Wit Nicolai theory \cite{deWit:1982ig}. In principle this uplift is known \cite{deWit:1986iy, Nicolai:2011cy} but in practice it is quite complicated. Simple formulae for this uplift are available in \cite{Cvetic1999b} for the STU-model when the axions vanish but our solutions have non-trivial axions and there is thus no simple uplift formula. In the symplectic frame discussed in \cite{Gnecchi:2013mta} the magnetic charges uplift to Chern numbers of fibration of $S^7$ over $\Sig_g$ (the structure group of the $S^7$ bundle is reduced from $SO(8)$ to $U(1)^4$) whereas in this frame the electric charges lift to angular momentum around circles in $S^7$. 

While our results are limited to BPS black holes, a solution for (not necessarily BPS) asymptotically AdS$_4$ black holes in the STU-model was presented in \cite{Chow:2013gba} depending on all eight charges. This was performed with a view towards embedding these black holes in the $\cN=8$ theory\footnote{see also \cite{Klemm:2012yg, Klemm:2012vm, Gnecchi:2012kb, Hristov:2013sya, Gnecchi:2014cqa, Barisch:2011ui,Goldstein:2014qha, Lu:2013ura} for other works on non-BPS black holes in $\cN=2$ gauged supergravity} using the embedding of \cite{Cvetic1999b}. In principle the solutions of the current work, when restricted to the STU-model should be contained in \cite{Chow:2013gba} but establishing this map does not appear to be straightforward. Presumably making this connection clean would aid our understanding of these solutions in general.

We hope that the solution obtained in this work will shine light on the construction of more general BPS black holes in gauged supergravity. One might find a useful a comparison to the work of Demianski-Plebanski \cite{Plebanski:1976gy} where they found a very general  black hole solution to the Einstein-Maxwell theory depending on mass, nut charge, electric and magnetic charge as well as rotation and acceleration, with and without a cosmological constant. This remarkable solution unified a slew of existing works in one framework (see \cite{Klemm:2013eca} for the most recent analysis of the supersymmetry of the PD solution). It seems to be a reasonable hope that a similar solution but with non-constant scalars can be obtained in FI-gauged supergravity, at least in the BPS sector, and we hope that the methods employed in this work, which maintain the full symmetries of the theory will be useful in this regard\footnote{See \cite{Kostelecky:1995ei, Chong:2004na, Klemm:2011xw, Chow:2013gba, Gnecchi:2013mja} for results on rotating black holes in gauged supergravity}. Having said that, one should keep in mind an interesting feature which emerged in the work \cite{Chow:2013gba}; while the full black hole solution space is continuously connected, allowing for a tuning to zero of the cosmological constant, the BPS sector is not. To pass continuously from a BPS asymptotically AdS$_4$ black hole to a BPS asymptotically flat black hole, one must pass through the non-BPS sector.

We also have in mind black holes in theories with hypermultiplets \cite{Halmagyi:2013sla}. Hypermultiplets do not add any charges, although they may give mass to vector fields, in a sense they are just an additional scalar field sector, and as such one might hope that ultimately there are analytic solutions analogous to those presented here with non-trivial profiles for the hyper-scalars. This would have interesting applications since the M-theory lift of the many known $\cN=2$ gauged supergravities with hypermultiplets often involve wrapped M5 branes in addition to wrapped M2-branes \cite{Donos:2012sy, Cassani:2012pj}. There are also known embeddings into M-theory of such theories which just contain the dynamics of M2-branes \cite{Bobev:2010ib}.

\vskip 1cm
%%%%%%%%%%%%%%%%%%%%%%%%%%%%%%%%%%%%%%%%%%%%%
\noindent {\bf Acknowledgments} I would like to thank Stefanos Katmadas for bringing his work to my attention and for discussions as well as Alessandra Gnecchi, Thomas Vanel and Harold Erbin for collaborations on closely related material. I offer my warm appreciation to the tranquility of the Parisian summer during which time these results were obtained. This work was conducted within the framework
of the ILP LABEX (ANR-10-LABX-63) supported by French state funds managed by the
ANR within the Investissements dÕAvenir programme under reference ANR-11-IDEX-0004-02
%%%%%%%%%%%%%%%%%%%%%%%%%%%%%%%%%%%%%%%%%%%%%

%%%%%%%%%%%%%%%%%%%%%%%%%%%%%%%%%%%%%%%%%%%%%
%%%%%%%%%%%%%%%%%%%%%%%%%%%%%%%%%%%%%%%%%%%%%
\begin{appendix}
%%%%%%%%%%%%%%%%%%%%%%%%%%%%%%%%%%%%%%%%%%%%%
%%%%%%%%%%%%%%%%%%%%%%%%%%%%%%%%%%%%%%%%%%%%%

%%%%%%%%%%%%%%%%%%%%%%%%%%%%%%%%%%%%%%%%%%%%%
\section{Special Geometry Background}\label{app:Special}
%%%%%%%%%%%%%%%%%%%%%%%%%%%%%%%%%%%%%%%%%%%%%
\subsection{Generalities}
%%%%%%%%%%%%%%%%%%%%%%%%%%%%%%%%%%%%%%%%%%%%%

We will use the conventions where the prepotential is given by
\be
\cF=-\frac{d_{ijk}X^i X^j X^k}{X^0}\,,
\ee
the metric is 
\be
g_{ij}=-\frac{3}{2}\frac{d_{y,ij}}{d_y} + \frac{9}{4}\frac{d_{y,i}d_{y,j}}{d^2_y}\,.
\ee
where
\be
X^\Lam=\bpm1 \\ z^i \epm\,,\quad\quad z^i=x^i+iy^i
\ee
and the covariant tensor is given by
\be
\hd^{ijk} = \frac{g^{il} g^{jm} g^{kn}d_{ijk}}{d_y^2}\,.
\ee
When $\cM_v$ is a homogeneous very special K\"ahler manifolds the tensor $\hd^{ijk}$ has constant entries and satisfies
\be
\hd^{ijk}d_{jl(m}d_{mp)k} =\frac{16}{27} \Bslb \delta^i_{l} d_{mnp} + 3 \delta^i_{(m} d_{mp)l} \Bsrb\,.
\ee
The sections are given by
\be
\cV=\bpm L^\Lam \\ M_\Lam \epm \ =\ e^{K/2} \bpm X^\Lam \\ F_\Lam  \epm
\ee
and satisfy\footnote{The symplectic inner product is $\langle A,B\rangle= B^\Lam A_\Lam-B_\Lam A^\Lam$.}
\be
\langle \cV,\cVbar \rangle = -i\,,\quad\quad \langle D_i\cV,D_{\jbar}\cVbar \rangle=ig_{i\jbar}
\ee
and any symplectic vector can be expanded in these sections. For example the charges are expanded as
\be
\cQ= i \cZbar \cV - i \cZ \cVbar + i \cZbar^{\ibar} D_{\ibar} \cVbar - i \cZbar^i D_i\cV
\ee
where 
\be
\cZ=\langle \cQ,\cV\rangle\,,\quad\quad \cZ_i = \langle \cQ,D_i \cV \rangle\,.
\ee
We also have a complex structure on the symplectic bundle over $\cM_v$:
\be
\Om\cM \cV= -i \cV\,,\quad\quad \Om\cM (D_i \cV)= i D_i \cV
\ee
where 
\be
\Om=\bpm0 & -1\!\!1 \\ 1\!\! 1 & 0 \epm\,,\quad\quad\cM=\bpm 1 & -\cR \\ 0 & 1 \epm\bpm \cI & 0 \\ 0 & \cI^{-1} \epm\bpm 1 & 0 \\ -\cR & 1 \epm
\ee
and $\cN=\cR+i\cI$ is the standard matrix which gives the kinetic and topological terms in the action for the gauge fields.

%%%%%%%%%%%%%%%%%%%%%%%%%%%%%%%%%%%%%%%%%%%%%
\subsection{The Quartic Invariant} \label{app:quartic}
%%%%%%%%%%%%%%%%%%%%%%%%%%%%%%%%%%%%%%%%%%%%%

Homogeneous very special K\"ahler manifolds were classified by de Wit and Van Proeyen \cite{deWit:1991nm} (see also \cite{deWit:1992wf, deWit:1993rr, deWit:1995tf}) and there are several infinite families as well as numerous sporadic examples. For each of these manifolds one can define the quartic invariant:
\bea
I_4(\cQ)&=& \frac{1}{4!}  t^{MNPQ}\cQ_M \cQ_N \cQ_P \cQ_Q \non \\
&=&-(p^0q_0+p^iq_i)^2 -4q_0 d_{ijk} p^ip^jp^k +\frac{1}{16}p^0 \hd^{ijk}q_iq_jq_k + \frac{9}{16} d_{ijk} \hd^{ilm} p^i p^kq_l q_m\,.
\eea
Recall that the indices take values $\Lam=0\,,\ldots\,,n_v$ and $i=1\,,\ldots\,, n_v$. Then the indices $\{M,N,P,Q\}$ take  both $\Lam$ indicies up and own, for example
\be
\cQ _M= \bpm p^\Lam \\ q_\Lam \epm\,.
\ee
Homogeneous spaces are cosets $G/H$ and $I_4(\cQ)$ is invariant under the global symmetries $G$ of the coset. In the work of de Wit and Van Proeyen one can find the explicit embedding of $G$ into the symplectic group $Sp(2n_v+2,\RR)$ which then acts in a straightforward manner on $I_4(\cQ)$. The first steps incorporating this quartic invariant into the lexicon of BPS black holes were taken in \cite{Cvetic:1995bj, Kallosh:1996uy, Cvetic:1996zq} it is quite remarkable how integral it has become. Some more recent references which utilize it are \cite{Cerchiai:2009pi, Ferrara:2011di, Bossard:2012xsa}.

Using the four index tensor $t^{MNPQ}$ one can also define $I_4$ evaluated on four distinct symplectic vectors as well as its derivative $I'_4$ which is itself a symplectic vector. We essentially use the same normalizations as in \cite{Katmadas:2014faa}:
\bea
I_4(A,B,C,D)&=& t^{MNPQ}A_M B_N C_P D_Q \\
I'_4(A,B,C)_M&=& \Om_{MN}t^{NPQR}A_N B_Q C_R \\
I_4(A)&=& \frac{1}{4!}  t^{MNPQ}A_M A_N A_P A_Q \\
I'_4(A)_M&=& \frac{1}{3!}  \Om_{MN}t^{NPQR}  A_P A_Q A_R
\eea
We then have
\be
24 I_4(A)= I_4(A,A,A,A)\,,\quad\quad 6 I'_4(A)= I'_4(A,A,A)\,.
\ee
A useful identity which plays the key role in deriving the form of the BPS equations given in \eq{BPSV1} and \eq{BPSV2} is
\be\label{I4pQImVImV}
I'_4(A,\Im\cV,\Im\cV)= 4 \Im\bslb \langle A,\cV \rangle\bsrb \Im \cV + 8 \Re\bslb \langle A,\cV \rangle\bsrb \Re \cV  - \Om \cM A\,.
\ee
Using this identity and replacing $A \ra \Im \cV$ we derive the useful expressions
\be\label{ReVImV}
\Re \cV = -\frac{I'_4(\Im\cV)}{2\sqrt{I_4(\Im\cV)}}\,.
\ee
and
\be
I_4(\Im\cV) = \frac{1}{16}\,.
\ee
%%%%%%%%%%%%%%%%%%%%%%%%%%%%%%%%%%%%%%%%%%%%%
\subsection{Identities Using the Quartic Invariant}
%%%%%%%%%%%%%%%%%%%%%%%%%%%%%%%%%%%%%%%%%%%%%
The components of $t^{MNPQ}$ are
\bea
&& t^{00}_{00}=-4 \,, \quad t^{0i}_{0j}=-2\delta^i_j \,, \quad t^{ij}_{kl}=-4\delta^{(i}_k \delta^{j)}_l+\frac{9}{4} d_{klm} \hd^{ijm} \,,  \\
&&\quad t^{ijk}_0=-\frac{3}{8}\hd^{ijk}\,,\quad t^0_{ijk}=24 d_{ijk}
\eea
from which with some effort one can derive the following identities. First we have the scalar identities
\bea
 \langle I'_4(\cG,\cG,\cQ),I'_4(\cG)\rangle &=& 8 I_4(\cG)\langle \cG,\cQ\rangle  \\
\langle I'_4(\cG,\cQ,\cQ),I'_4(\cG)\rangle&=& \frac{2}{3} I_{4}(\cQ,\cG,\cG,\cG) \langle \cG,\cQ\rangle    \\
 \langle I'_4(\cG,\cQ,\cQ),I'_4(\cQ,\cG,\cG)\rangle &=& 4 I_4(\cQ,\cQ,\cG,\cG) \langle \cG,\cQ\rangle-12 \langle I'_4(\cQ),I'_4(\cG)\rangle   \,.
\eea
Then  to evaluate the LHS of \eq{BPSorder1}--\eq{BPSorder5} we need to expand various symplectic vectors in terms of the set
\be
\{ \cG,\cQ,I_4'(\cG),I_4'(\cG,\cG,\cQ),I_4'(\cG,\cQ,\cQ),I_4'(\cQ) \}\,.
\ee
After much algebra we found the following expressions:
\bea
I_4'(I'_4(\cG),I_4'(\cG),\cG)&=&8 I_4(\cG) I_4'(\cG) \non \\
\non \\%
I_4'(I'_4(\cG),I_4'(\cG,\cG,\cQ),\cG)&=&2  I_4(\cG,\cG,\cG,\cQ) I_4'(\cG)+ 18 I_4(\cG) \langle \cG,\cQ \rangle \cG  \non  \\
\non \\%
I_4'(I'_4(\cG),I_4'(\cG,\cQ,\cQ),\cG)&=& \frac{4}{3} I_4(\cG,\cG,\cG,\cQ)  \langle \cG,\cQ \rangle \cG+2 I_4( \cG,\cG,\cQ,\cQ)I'_4(\cG) \non \\
\non \\%
I_4'(I'_4(\cG),I'_4(\cQ),\cG)&=& 2  \langle I'_4(\cG),I'_4(\cQ) \rangle \cG+\frac{1}{3} I_4(\cG,\cQ,\cQ,\cQ) I'_4(\cG)  \non
\eea
\bea
I_4'(I'_4(\cG,\cG,\cQ),I'_4(\cG,\cG,\cQ),\cG)&=& 8 I_4(\cG,\cG,\cQ,\cQ) I'_4(\cG) +\frac{4}{3} I_4(\cG,\cG,\cG,\cQ)I'_4(\cG,\cG,\cQ) \non  \\
&& -16 I_4(\cG) I'_4(\cG,\cQ,\cQ) + 64 I_4(\cG) \langle \cG,\cQ \rangle \cQ \non \\
&&+\frac{16}{3} I_4(\cG,\cG,\cG,\cQ) \langle \cG,\cQ \rangle \cG \non \\
\non \\%
I_4'(I_4(\cG,\cG,\cQ),I_4'(\cG,\cQ,\cQ),\cG)&=& \frac{16}{3} I_4(\cG,\cQ,\cQ,\cQ)I'_4(\cG) +2 I_4(\cG,\cG,\cQ,\cQ) I'_4(\cG,\cG,\cQ)\non \\
&& -\frac{2}{3} I_4(\cG,\cG,\cG,\cQ)I'_4(\cG,\cQ,\cQ)  -32 I_4(\cG) I'_4(\cQ) \non \\
&& +8\Bslb  I_4(\cG,\cG,\cQ,\cQ)\langle \cG,\cQ \rangle - \langle I'_4(\cG), I'_4(\cQ) \rangle\Bsrb \cG \non \\ 
&&+\frac{16}{3} I_4(\cG,\cG,\cG,\cQ)\langle \cG,\cQ \rangle \cQ\non \\
\non \\%
I_4'(I'_4(\cG,\cG,\cQ),I'_4(\cQ),\cG)&=& -\frac{2}{3}  I_4(\cG,\cG,\cG,\cQ)I'_4(\cQ)+ 8 \langle I'_4(\cG),I'_4(\cQ) \rangle \cQ  \non \\
&&+\frac{4}{3} I_4(\cG,\cQ,\cQ,\cQ)\langle \cG,\cQ \rangle \cG +16 I_4(\cQ) I'_4(\cG) \non  \\
&& + \frac{1}{3} I_4(\cG,\cQ,\cQ,\cQ) I'_4(\cG,\cG,\cQ) \non 
\eea
\bea
I_4'(I'_4(\cG,\cQ,\cQ),I_4'(\cG,\cQ,\cQ),\cG)&=& 32 I_4(\cQ) I'_4(\cG) + \frac{8}{3} I_4(\cG,\cQ,\cQ,\cQ) I'_4(\cG,\cG,\cQ)  \non \\ 
&&- \frac{16}{3} I_4(\cG,\cG,\cG,\cQ) I'_4(\cQ) +16 I_4(\cG,\cG,\cQ,\cQ) \langle \cG,\cQ\rangle \cQ \non \\
&& +\frac{16}{3} I_4(\cG,\cQ,\cQ,\cQ) \langle \cG,\cQ\rangle \cG -32 \langle I'_4(\cG),I'_4(\cQ) \rangle  \cQ \non \\
\non \\%
I_4'(I'_4(\cG,\cQ,\cQ),I_4'(\cQ),\cG)&=& 8 I_4(\cQ) I'_4(\cG,\cG,\cQ) + \frac{1}{3} I_4(\cG,\cQ,\cQ,\cQ) I'_4(\cG,\cQ,\cQ) \non \\
&& -2 I_4(\cG,\cG,\cQ,\cQ) I'_4(\cQ) + \frac{8}{3} I_4(\cG,\cQ,\cQ,\cQ) \langle \cG,\cQ \rangle \cQ \non\\
&& +16  I_4(\cQ) \langle \cG,\cQ \rangle \cG \non \\
\non \\%
I_4'(I'_4(\cQ),I_4'(\cQ),\cG)&=&  4 I_4(\cQ) I'_4(\cG,\cQ,\cQ)  - \frac{2}{3} I_4(\cG,\cQ,\cQ,\cQ) I'_4(\cQ)\non\\
&&+ 16  I_4(\cQ) \langle \cG,\cQ \rangle \cQ  \non 
\eea

%%%%%%%%%%%%%%%%%%%%%%%%%%%%%%%%%%%%%%%%%%%%%
\end{appendix}
%%%%%%%%%%%%%%%%%%%%%%%%%%%%%%%%%%%%%%%%%%%%%

%%%%%%%%%%%%%%%%%%%%%%%%%%%%%%%%%%%%%%%%%%%%%
%%%%%%%%%%%%%%%%%%%%%%%%%%%%%%%%%%%%%%%%%%%%%
\providecommand{\href}[2]{#2}\begingroup\raggedright\endgroup

%%%%%%%%%%%%%%%%%%%%%%%%%%%%%%%%%%%%%%%%%%%%%
%%%%%%%%%%%%%%%%%%%%%%%%%%%%%%%%%%%%%%%%%%%%%

%%%%%%%%%%%%%%%%%%%%%%%%%%%%%%%%%%%%%%%%%%%%%
\end{document}